\documentclass[12pt]{book}
\usepackage{myplenum,epsf}
\begin{document}

\chapter{CLASSICAL BEHAVIOR OF A MACROSCOPIC SCHR\"ODINGER CAT}

\author{Carlo Presilla}

\affiliation{Dipartimento di Fisica, 
Universit\`a di Roma ``La Sapienza''\\
Piazzale A. Moro 2, Rome, Italy, 00185\\
and INFM, Unit\`a di Ricerca di Roma ``La Sapienza''}

\abstract{
We study the dynamics of classical and quantum systems linearly 
interacting with a classical environment represented by an 
infinite set of harmonic oscillators. 
The environment induces a dynamical localization of the quantum 
state into a generalized coherent state for which the 
$\hbar \to 0$ limit always exists and reproduces the classical motion.
We describe the consequences of this localization on the behavior
of a macroscopic system by considering the example of a
Schr\"odinger cat.}

\section{INTRODUCTION}
The problem of how classical behavior is regained from quantum 
mechanics in the macroscopic limit can be conceptually solved by 
recognizing that a macroscopic system is never completely 
isolated by the external world. 
It has been argued\refnote{\cite{JOOSZEH,ZUREK,CINI}} 
that the interaction with an environment can, 
after a transient whose duration presumably depends on the 
coupling strength, drive the totality of the admissible states 
of the Hilbert space into those having classical limit,
formally $\hbar \to 0$. 

In a recent paper in collaboration with R. Onofrio and M. Patriarca\refnote{\cite{POP}}, we substantiated this conjecture  
by analyzing the dynamics of general classical and 
quantum systems linearly interacting with an infinite set of 
degrees of freedom.
Here, we briefly review the main results of this model and 
describe in detail how the pathologies of a simple Schr\"odinger 
cat are cured by the presence of the environment. 

\section{DYNAMICS OF SYSTEMS INTERACTING WITH AN ENVIRONMENT}
Let us consider a system described by the classical Hamiltonian
\begin{equation} 
H(p,q,t) = {p^2 \over {2m}}+V(q,t)
\label{H}
\end{equation}
We model its interaction with a classical environment by a 
linear coupling to an infinite set of degrees of freedom
$\{P_n, Q_n\}$.
The Hamiltonian for the total system is 
\begin{equation} 
H_{tot} = H(p,q,t) + H_m(P,Q-q), 
\label{HTOT}
\end{equation}
where
\begin{equation}
H_m(P,Q-q) = \sum_n \left[ {P_n^2 \over {2M}}+
{M \omega_n^2 \over 2} (Q_n-q)^2 \right].
\label{HM}
\end{equation}

The classical dynamics of the system modified by the environment
is described in terms of equations obtained by formally solving 
the harmonic motion of $\{P_n, Q_n\}$.
For an environment having frequencies $\{ \omega_n \}$ distributed
with density 
$dN / d\omega=\theta(\Omega-\omega) 2 m \gamma / \pi M \omega^2$,
we get, for times $\gg \Omega^{-1}$, the Markovian evolution 
\begin{eqnarray}
dp(t) &=& - \left[ \gamma p(t) + \partial_q V(q(t),t) \right] dt 
+ \sqrt{ 2 m \gamma k_B T} \xi(t)dt
\label{LANGE1} \\
dq(t) &=& {p(t) \over m} dt
\label{LANGE2}
\end{eqnarray}
with initial conditions $p'$ and $q'$ at time $t'$. 
If the the initial conditions $\{ P_n', Q_n' \}$ of the 
environment are chosen
as a realization of the equilibrium Gibbs measure at temperature
$T$, then $\xi(t)$ is a realization of a stochastic process
in time with respect to the same measure with properties
$\overline{\xi(t)} =0$ and $\overline{\xi(t) \xi(s)}=\delta(t-s)$.
Therefore, Eqs. (\ref{LANGE1},\ref{LANGE2}) are stochastic 
Langevin equations.
 
In alternative to the detailed stochastic description,
we may be interested to determine the average behavior
of the system obtained by considering all the possible 
realizations of the initial conditions of the environment.
In this case, the system is described by a probability density 
$W(p,q,t)$ solution of the Fokker-Plank equation associated to 
(\ref{LANGE1}-\ref{LANGE2}) 
\begin{equation}
\partial_t W(p,q,t) = 
\left[ - {p \over m} \partial_q + \partial_q V(q,t) \partial_p 
+ \partial_p \left( \gamma p + m \gamma k_BT \partial_p \right)
\right] W(p,q,t)
\label{FOKKERPLANK}
\end{equation}
with initial conditions $W(p,q,t')= \delta(p-p') \delta(q-q')$.

The classical analysis can be repeated at quantum level.
Besides obvious technical modifications, there is now
a conceptual difference.  
Since we do not know how to describe the coupling of
classical and quantum degrees of freedom, we must start with 
a quantum description of both the system and the environment. 
The condition of classical behavior of the environment can be 
reintroduced later by asking that the thermal energy 
$k_BT$ is much larger than the energy spacing of the 
highest-frequency oscillators $\hbar \Omega$.
If this high temperature condition is satisfied, for times 
$\gg \Omega^{-1}$ the system is described by the nonlinear 
stochastic Schr\"odinger equation 
\begin{eqnarray}
d | \psi_{[\xi]}(t) \rangle &=& 
- {i \over \hbar} \left[ \hat H(\hat p,\hat q,t) +
{\gamma \over 4} (\hat{p}\hat{q}+\hat{q}\hat{p}) \right]
| \psi_{[\xi]}(t) \rangle dt
\nonumber \\ &&
-{1\over2} \left[ \hat{A}^{\dag}\hat{A} + a(t)^*a(t) - 
2a(t)^*\hat{A}\right] | \psi_{[\xi]}(t) \rangle dt
+\left[ \hat{A} - a(t) \right] | \psi_{[\xi]}(t) \rangle 
\xi(t)dt,~~~~~
\label{PSTOCEQ}
\end{eqnarray}
where
$\hat{A}=\sqrt{2m \gamma k_BT / \hbar^2} ~\hat{q} + 
i \sqrt{\gamma / 8 m k_BT} ~\hat{p}$,
$a(t)=\langle \psi_{[\xi]}(t)| \hat{A} |\psi_{[\xi]}(t) \rangle$,
and $\xi(t)$ is a real white noise.

A direct characterization of the average properties
of the quantum system is also possible.
By introducing the reduced density matrix operator
\begin{equation}
\hat{\rho}(t) = 
\overline{|\psi_{[\xi]}(t) \rangle \langle \psi_{[\xi]}(t)|}
\label{PSIPSISUM}
\end{equation}
and the associated Wigner function 
\begin{equation}
W(p,q,t)= {1 \over 2 \pi \hbar} 
\int dz~ \exp \left({i\over\hbar} pz \right)
\langle q - {z\over 2}| \hat{\varrho}(t) |q+ {z\over 2} \rangle,
\end{equation}
from Eq. (\ref{PSTOCEQ}) we obtain
\begin{eqnarray}
\partial_t W(p,q,t) = \biggl[ 
&-&{p \over m} \partial_q 
+\sum_{n=0}^{\infty} \left({\hbar \over 2i} \right)^{2n}
{1 \over (2n+1)!}\partial_q^{2n+1} V(q,t) \partial_p^{2n+1} 
\nonumber \\ &+&
\partial_p \left( \gamma p + m \gamma k_BT \partial_p \right) 
+ {\hbar^2 \gamma \over 16mk_BT} \partial_q^2 \biggr] W(p,q,t).
\label{PCLWEQ}
\end{eqnarray}
Note that (\ref{PSTOCEQ}) and (\ref{PCLWEQ}) reduce to the
corresponding quantum equations for an isolated system 
when $\gamma=0$.

\section{DYNAMICAL LOCALIZATION INTO A COHERENT STATE}
In the previous section, we have described the equations which
govern the dynamics of classical and quantum systems in 
interaction with a classical environment.
Now, we show that the quantum dynamics reduces
to the classical one when the limit $\hbar \to 0$ is taken.
Note that this is not always possible in the case of an 
isolated system where well known pathological limits
are encountered\refnote{\cite{SCHROEDINGER}}.

It is possible to demonstrate 
rigorously\refnote{\cite{HALLIWELL96},\cite{POP}} that 
for a system with potential
\begin{equation} 
V(q,t)=v_0(t)+v_1(t)q+\frac{1}{2}m\omega_0^2 q^2
\label{QUADPOT}
\end{equation}
after a time which, in the worst case, is of the order of
$\sqrt{\hbar / \gamma k_BT}$ the solutions of Eq. (\ref{PSTOCEQ}) 
become of the form  
\begin{equation}
|\psi_{[\xi]}(t)\rangle = 
\exp\left[-{i \over \hbar} \varphi(t) \right]~|p(t)q(t)\rangle,
\end{equation}
where $\varphi(t)$ is real,
$p(t)=\langle \psi_{[\xi]}(t)|\hat{p}| \psi_{[\xi]}(t) \rangle$, 
$q(t)=\langle \psi_{[\xi]}(t)|\hat{q}| \psi_{[\xi]}(t) \rangle$, 
and  
\begin{eqnarray}
\langle q |p(t)q(t)\rangle &=& (2\pi\sigma_q^2)^{-1/4}
\exp \left\{ - {1 - {2i\over \hbar} \sigma_{pq}^2 \over 
4 \sigma_q^2}~[q-q(t)]^2 + {i\over \hbar} p(t)[q-q(t)] \right\}.
\label{LTPSI}
\end{eqnarray}
The state (\ref{LTPSI}) defined in terms of the parameters
\begin{eqnarray}
\sigma_q^2 &=& {\hbar \over m} \sqrt{\gamma^2 -\omega_0^2 + 
\sqrt{(\gamma^2-\omega_0^2)^2 + 
16 \left(\gamma k_BT/\hbar \right)^2}
\over 32 \left(\gamma k_BT/\hbar \right)^2 } 
\label{SIGMAQSOL} \\
\sigma_{pq}^2 &=& \sqrt{m^2(\gamma^2-\omega_0^2) \sigma_q^4 + 
{\hbar^2\over 4}} - m\gamma \sigma_q^2
\label{SIGMAPQSOL}
\end{eqnarray}
is a generalized coherent state which admits the $\hbar \to 0$
limit. 
Moreover, the convergence into this coherent state takes place
in a time which vanishes for $\hbar \to 0$.
Therefore, a linear system like (\ref{QUADPOT}) always has 
classical limit at any time $t>t'$ even if the $\hbar \to 0$ limit
does not exist at the initial time $t'$. 
Since the contribution to the Green function of Eq. 
(\ref{PSTOCEQ}) due the system-environment coupling
is of the form $q^2 \hbar^{-1}$, these results
apply in the limit $\hbar \to 0$ also to nonlinear systems.

We illustrate the consequences of the localization into a
coherent state induced by the coupling with the environment
by analyzing the evolution of a cat state.   
For simplicity, consider a free quantum particle which at 
time $t'=0$ is in the superposition state
\begin{equation}
|\psi(0) \rangle = N 
\left( |\frac{1}{2}P -\frac{1}{2}Q\rangle + 
|-\frac{1}{2}P \frac{1}{2}Q \rangle \right) 
\label{CAT}
\end{equation}
where $|\pm \frac{1}{2} P \pm\frac{1}{2}Q\rangle$
are coherent states (\ref{LTPSI}-\ref{SIGMAPQSOL}) 
with $\omega_0=0$ and $N$ is a normalization factor.
The state (\ref{CAT}) has no classical counterpart.
Indeed, its corresponding Wigner function
\begin{eqnarray}
W(p,q,0) = N^2 &\Bigg\{& 
W_{\frac{1}{2}P-\frac{1}{2}Q}(p,q)+
W_{-\frac{1}{2}P\frac{1}{2}Q}(p,q) 
+W_{00}(p,q)~
2 \cos \Bigg[ {p \over \hbar}Q + {q \over \hbar} P \Bigg]  
\Bigg\}~~~~~
\label{WCAT0}
\end{eqnarray} 
does not have $\hbar \to 0$ limit due to the presence of the
last oscillating term.
In Eq. (\ref{WCAT0}), we indicated with $W_{p_0q_0}(p,q)$ 
the Wigner function of the coherent state $|p_0q_0 \rangle$ 
having limit
$W_{p_0q_0}(p,q)\stackrel{\hbar \to 0}{\longrightarrow}
\delta(p-p_0)\delta(q-q_0)$.
In more physical terms, if we consider a macroscopic limit, 
for instance increase the values of $P$ and/or $Q$, 
the oscillating interference term in (\ref{WCAT0}) 
never disappear contrarily to common sense. 
An example of this pathological behavior is shown in Fig. 1.
\begin{figure}[htb]
\vskip -1cm
\begin{center}
\epsfxsize=.8\hsize
\leavevmode\epsffile{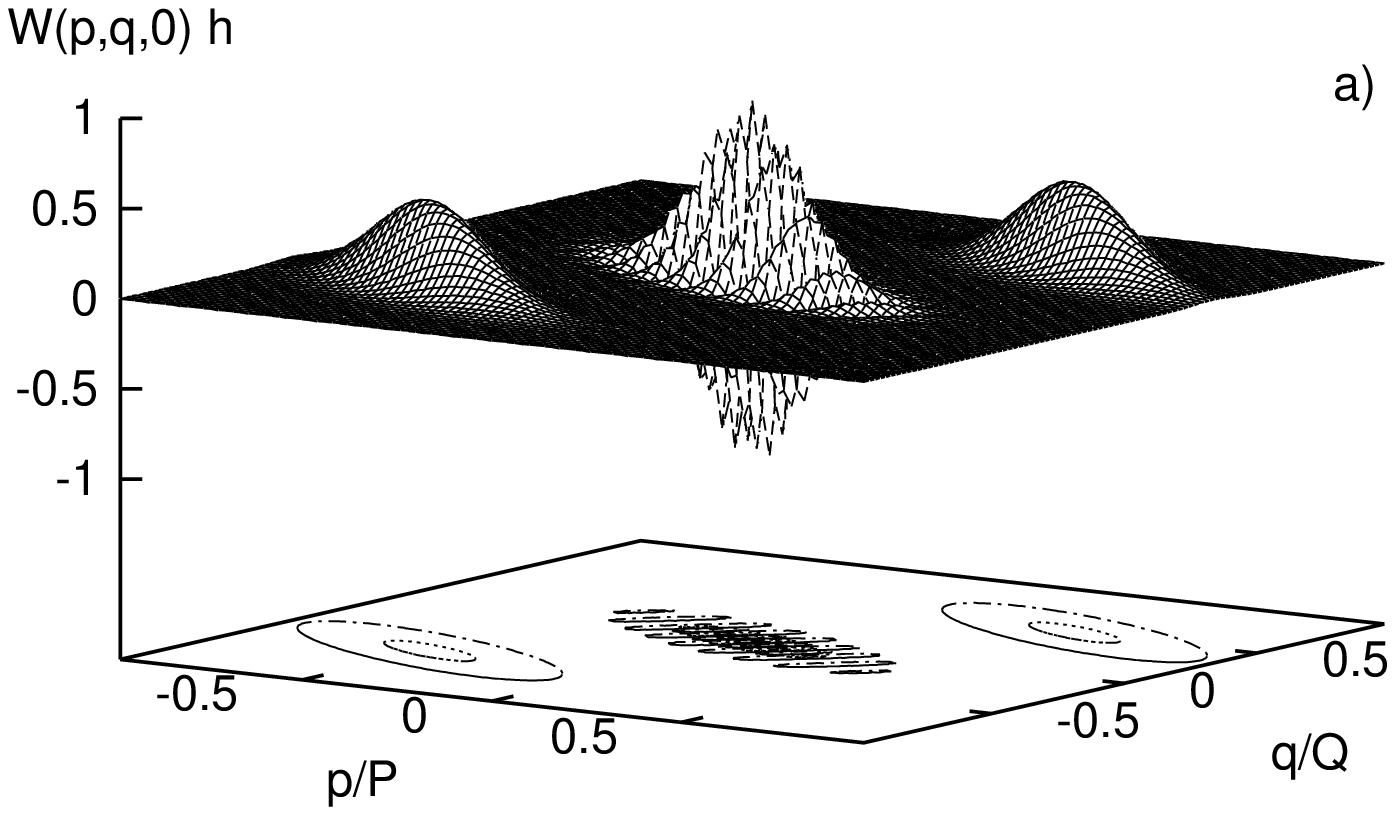}
\end{center}
\begin{center}
\epsfxsize=.8\hsize
\leavevmode\epsffile{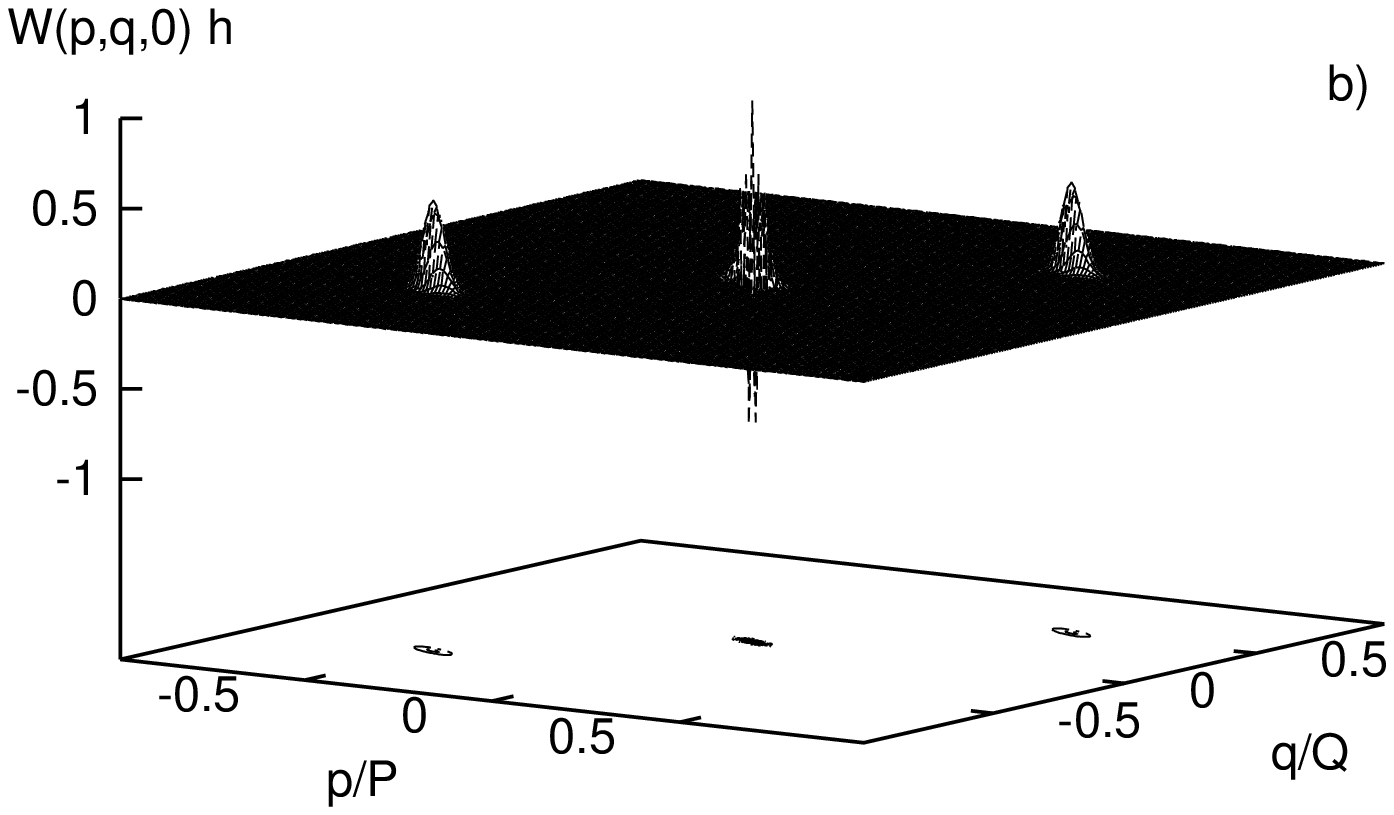}
\end{center}
\vskip -1cm
\caption{Wigner function at time 0 corresponding to the cat state
(\protect{\ref{CAT}}) for $k_BT / \hbar \gamma=100$.
In case a) we have $P=20 \sqrt{\hbar m \gamma}$ 
and $Q=2 \sqrt{m \gamma / \hbar}$
while in b) $P=100 \sqrt{\hbar m \gamma}$ 
and $Q=10 \sqrt{m \gamma / \hbar}$.}
\end{figure}

At a later time, the situation is different.
By solving Eq. (\ref{PCLWEQ}) with the initial condition 
(\ref{WCAT0}), we find the following expression for the Wigner 
function  
\begin{eqnarray}
W(p,q,t) &=& N^2 \Bigg\{ 
W_{\frac{1}{2}P-\frac{1}{2}Q}(p,q,t)+
W_{-\frac{1}{2}P+\frac{1}{2}Q}(p,q,t) 
+W_{00}(p,q,t)
\nonumber \\ && \times
e^{-C_{P-Q}+ \Sigma_{P-Q}(t)}
~2\cos \left[ 
p \Upsilon_{P-Q}(t) + q \Phi_{P-Q}(t)  \right]  \Bigg\}.
\label{WCAT}
\end{eqnarray} 
The definitions of $W_{p_0q_0}(p,q,t)$, $C_{P-Q}$, 
$\Sigma_{P-Q}(t)$, $\Upsilon_{P-Q}(t)$, and 
$\Phi_{P-Q}(t)$ can be found in Ref. \cite{POP}. 
The functions $W_{p_0q_0}(p,q,t)$ are solution of Eq. 
(\ref{PCLWEQ}) with initial condition $W_{p_0q_0}(p,q)$.
In the $\hbar \to 0$ limit, they reduce to phase-space 
probability densities $W^{\rm cl}_{p_0q_0}(p,q,t)$ 
solution of the classical Fokker-Plank equation 
(\ref{FOKKERPLANK}) with initial condition 
$\delta(p-p_0) \delta(q-q_0)$.
The exponential term $\exp[-C_{P-Q}+\Sigma_{P-Q}(t)]$ 
vanishes for $\hbar \to 0$ at any $t>0$ and 
\begin{equation}
\lim_{\hbar \to 0} W(p,q,t) = 
{1\over 2} \left[ 
W_{\frac{1}{2}P-\frac{1}{2}Q}^{\rm{cl}}(p,q,t)+
W_{-\frac{1}{2}P\frac{1}{2}Q}^{\rm{cl}}(p,q,t) \right].
\label{WCATCL}
\end{equation}
The classical limit is equivalently reached for large 
values of $P$ and/or $Q$.
An estimate of the critical values of $P$ and $Q$ 
for which the Wigner function changes from (\ref{WCAT})
to (\ref{WCATCL}) can be obtained by observing that 
$\Sigma_{P-Q}(t) = C_{P-Q} -t/\tau + {\cal O}(t^2)$
with the characteristic time $\tau$ given by 
\begin{equation}
\gamma \tau \simeq \left(
{k_BT \over \hbar \gamma}~{\hbar Q^2 \over m \gamma}
+ {1 \over 16} {\hbar \gamma \over k_BT}~
{P^2 \over \hbar m \gamma} \right)^{-1}.
\label{GAMMATAU}
\end{equation}
Note that $\tau(P,Q) \to 0$ when $P$ and/or $Q$ diverge.
At a chosen time $t>0$, we have quantum behavior for 
$\tau(P,Q) \gg t$ (microscopic system) and classical behavior 
for $\tau(P,Q) \ll t$ (macroscopic system).
An example of this quantum-to-classical transition is shown 
in Fig. 2. 
\begin{figure}[htb]
\vskip -1cm
\begin{center}
\epsfxsize=.8\hsize
\leavevmode\epsffile{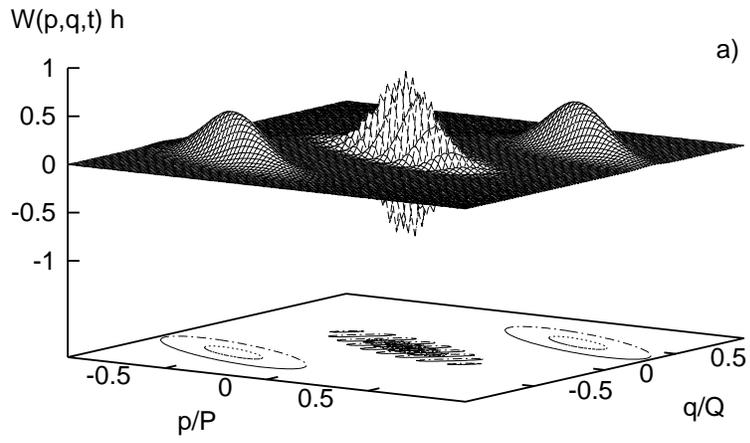}
\end{center}
\begin{center}
\epsfxsize=.8\hsize
\leavevmode\epsffile{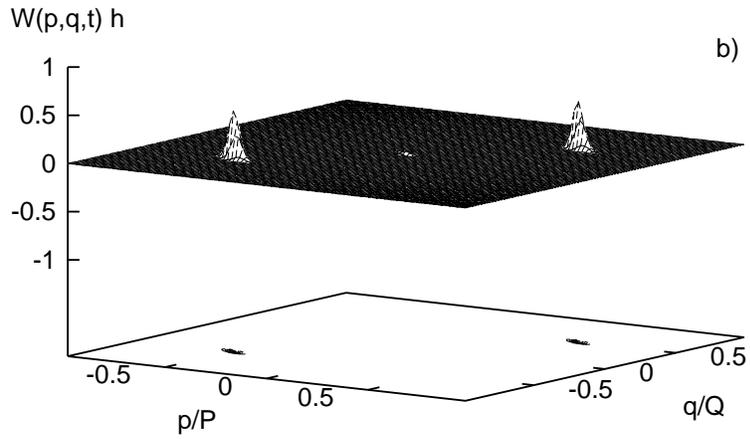}
\end{center}
\vskip -1cm
\caption{As in Fig. 1 but at the adimensional time 
$\gamma t= \frac{1}{2} 10^{-3}$. 
Note that the adimensional characteristic time 
(\protect{\ref{GAMMATAU}}) is about $\frac{1}{4} 10^{-2}$ 
in case a) and $10^{-4}$ in case b).}
\end{figure}

Finally, we note that $\Sigma_{P-Q}(t)$, $\Upsilon_{P-Q}(t)$,
and  $\Phi_{P-Q}(t)$ vanish for $t \to \infty$ so that in this 
limit the Wigner function becomes
\begin{eqnarray}
W_\infty(p,q,t) = N^2 &\Bigg\{& 
W_{\frac{1}{2}P-\frac{1}{2}Q}^{\rm{cl}}(p,q,t)+
W_{-\frac{1}{2}P\frac{1}{2}Q}^{\rm{cl}}(p,q,t) 
+W_{00}^{\rm{cl}}(p,q,t) e^{-C_{P-Q}}2 \Bigg\}.
\label{WCATLT}
\end{eqnarray} 
The classical limit, formally $\hbar \to 0$, is reached only 
for a macroscopic system, not in the long time limit of a 
microscopic one.

\end{document}